\documentclass[nohyperref]{article}
\usepackage{microtype}
\usepackage{graphicx}
\usepackage{booktabs} 

\usepackage{hyperref}

\usepackage[accepted]{icml2023}

\usepackage{amsmath}
\usepackage{amssymb}
\usepackage{mathtools}
\usepackage{amsthm}

\usepackage[utf8]{inputenc} 
\usepackage[T1]{fontenc}    
\usepackage{url}            
\usepackage{booktabs}
\usepackage{multirow}
\usepackage{amsfonts}       
\usepackage{nicefrac}       
\usepackage{xcolor}         
\usepackage{pgfplots}
\usepackage{pgf}
\usepackage{array}
\usepackage{overpic}
\usepackage{wrapfig}
\usepackage{caption}
\usepackage{subcaption}
\usepackage[capitalize,noabbrev]{cleveref}

\theoremstyle{plain}

\theoremstyle{definition}

\theoremstyle{remark}

\usepackage[textsize=tiny]{todonotes}

\icmltitlerunning{GEP}

\begin{document}
\twocolumn[
\icmltitle{Geometric Epitope and Paratope Prediction}


\icmlsetsymbol{equal}{*}

\begin{icmlauthorlist}
\icmlauthor{Marco Pegoraro}{equal,sap}
\icmlauthor{Clémentine Dominé}{equal,gatsby}
\icmlauthor{Emanuele Rodolà}{sap}
\icmlauthor{Petar Veli{\v{c}}kovi{\'c}}{comp}
\icmlauthor{Andreea Deac}{mila}
\end{icmlauthorlist}

\icmlaffiliation{sap}{Sapienza, University of Rome}
\icmlaffiliation{gatsby}{Gatsby Computational Neuroscience Unit, University College London}
\icmlaffiliation{comp}{Google DeepMind}
\icmlaffiliation{mila}{Mila, Université de Montréal}

\icmlcorrespondingauthor{Marco Pegoraro}{pegoraro@di.uniroma1.it}
\icmlcorrespondingauthor{Clémentine Dominé}{clementine.domine.20@ucl.ac.uk}

\icmlkeywords{Machine Learning, ICML}

\vskip 0.3in
]
\printAffiliationsAndNotice{\icmlEqualContribution}

\begin{abstract}
  Antibody-antigen interactions play a crucial role in identifying and neutralizing harmful foreign molecules. In this paper, we investigate the optimal representation for predicting the binding sites in the two molecules and emphasize the importance of geometric information. Specifically, we compare different geometric deep learning methods applied to proteins' inner (I-GEP) and outer (O-GEP) structures. We incorporate 3D coordinates and spectral geometric descriptors as input features to fully leverage the geometric information. 
Our research suggests that surface-based models are more efficient than other methods, and our O-GEP experiments have achieved state-of-the-art results with significant performance improvements.
\end{abstract}

\section{Introduction}

Identifying the binding sites of antibodies is essential for developing vaccines and synthetic antibodies. These binding sites, called paratopes, can bind to antigens, wherein the corresponding binding site is known as the epitope, thus neutralizing harmful foreign molecules in the body.  
Experimental methods for determining the residues that belong to the paratope and epitope are time-consuming and expensive, highlighting the need for computational tools to facilitate the rapid development of therapeutics. The recent COVID-19 epidemic highlighted this need further, as mutations in the antigen were shown to impact the binding mechanism, potentially reducing the efficacy of existing treatments \cite{thomson2021circulating}. Predicting the binding sites of an antibody-antigen interaction requires considering the entire antigen for epitope prediction and a localized region of the antibody, known as the Complementarity-Determining Region (CDR), for paratope prediction. 

The integration of geometric and structural information in protein-to-protein interaction studies has led to significant progress \cite{stark2022equibind, PiNet}.  While several methods have concentrated on the 3D graph representation, few methods \cite{PiNet, zhang2023equipocket} have investigated the 3D surface representation. We aim to assess the impact of utilizing the geometric representation of the antigen and antibody in the task of epitope-paratope prediction.
Our approach, GEP (Geometric Epitope-Paratope) Prediction, proposes different geometric representations of the molecules to create accurate predictors for predicting antibody-antigen binding sites.
The use of geometrical information is further justified by the emergence of technology predicting the single-protein structure, such as AlphaFold 2 \cite{jumper2021highly}, which has comparable accuracy to experimental methods. 
 We present the following contributions in our paper:
\begin{itemize}
\item We analyze the significance of geometric information within the context of graph learning, using equivariant layers that enable more robust and accurate predictions. 
\item Additionally, we fully exploit the geometric information in molecules by representing them as surfaces and applying techniques based on spectral geometry, leading to state-of-the-art performance.
\item We will release a pipeline for generating a dataset from PDB molecules that produces molecular representations in graph and surface formats, enabling cross-method comparisons.
\end{itemize}
\section{Related work}

The structure of proteins provides crucial information about the location and orientation of the binding sites. Various approaches have been taken in the literature to address the task of epitope and paratope prediction, including sequential \cite{liberis2018parapred,deac2019fastparapred} and structural \cite{krawczyk2014improving,del2021neural} methods. 
Furthermore, Geometric deep learning has emerged as a powerful tool for predicting protein-protein interactions \cite{isert2023structure}, with graph-based representations being one of the most common approaches \cite{tubiana2022scannet,stark2022equibind}. These methods leverage the geometric information of the molecules to learn complex relationships between epitopes and paratopes. For instance, some approaches \cite{del2021neural,da2022epitope3d} use the graph structure to compute features based on neighbouring residues, which are then aggregated to highlight the most probable region of interaction.

An alternative approach is to represent proteins as surfaces. 
MaSIF \cite{gainza2020MaSif} focuses on the more general problem of protein interaction region prediction and uses a surface representation learned through convolutions defined on the surface.
PiNet \cite{PiNet} represents the protein surface as a point cloud and employs PointNet \cite{qi2017pointnet} to classify points as interacting or not. On the contrary, \citet{zhang2023equipocket} model the surface of a molecule as a graph and apply an equivariant graph neural network (EGNN, \cite{satorras2021n}) for binding site prediction. 

Integrating structural and geometric information has proven to be a promising approach for improving protein interaction prediction. Still, few studies have focused on the specific case of epitope and paratope prediction \cite{cia2023critical}. Our work supports this view by showing that considering the problem as a geometric one can effectively improve performance.

\section{Motivation}
The shape and structure of molecules play a crucial role in determining their interactions with other molecules, as complementary geometric shapes are required for successful binding \cite{fischer1894einfluss}. To accurately predict molecular interactions, it is essential to incorporate geometric information such as 3D coordinates and spectral descriptors. Our approach to predicting molecular interactions integrates this geometric information into the representation of proteins as graph residues, resulting in a more enhanced and accurate representation.Furthermore, we recognize the importance of the outer surface of a molecule in molecular interactions. To address this, we focus on computations performed on the outer surface of the molecule and then map these predictions to the corresponding residues. By considering the surface of the molecule, we gain valuable insights into the molecular interactions occurring on the surface and enable the use of geometric deep-learning models to analyze these interactions. This approach can potentially provide significant benefits over traditional methods, ultimately leading to more accurate and efficient predictions of molecular interactions.

\section{Data}

Comparing methods across different molecular representations is crucial for advancing research in molecular modelling. We developed a reusable pipeline that generates a dataset to evaluate methods using inner and outer structure representations. 

We collected a dataset of 133 protein complexes from Epipred \cite{krawczyk2014improving}, with 103 for training and 30 for testing. The training and test sets have been selected to share no more
than 90\% pairwise sequence identity. The PDB files were obtained from the Sabdab database \cite{dunbar2014sabdab}. In the test set, 7.8\% of antigen residues were labelled as positive. Additionally, we used a separate set of 27 protein complexes from PECAN derived from a subset of the Docking Benchmark v5 \cite{vreven2015updates} to validate our results.


We construct a residue graph (Figure \ref{fig:graph_res}) for each protein, where a 28-dimensional physicochemical feature vector represents each residue. This vector comprises a one-hot encoding of the amino acid (including 20 possible types and one for an unknown type), in addition to seven other features representing the physical, chemical, and structural properties of the amino acid type. These additional features can be considered a fixed embedding, as described in \cite{meiler2001generation}.

For each protein, we generated a surface mesh (Figure \ref{fig:surf_res}) using the PyMOL API with a 1.4~\AA~water probe radius. We associated each point on the protein's surface with a residue by finding the closest atom to that point. 
This association was then used to transfer the feature of each residue to the points on the surface.

\section{Method}
In our experiments, we considered two scenarios: a protein represented through its inner structure (\textit{I-GEP}) and outer structure (\textit{O-GEP}). In both cases, we leverage the geometric information to improve the performance of epitope and paratope prediction methods.

\begin{figure}[h]
\vspace{0cm}
\centering
\begin{overpic}
[trim=0cm 0cm 0cm 0cm,clip,width=0.99\linewidth, grid=false]{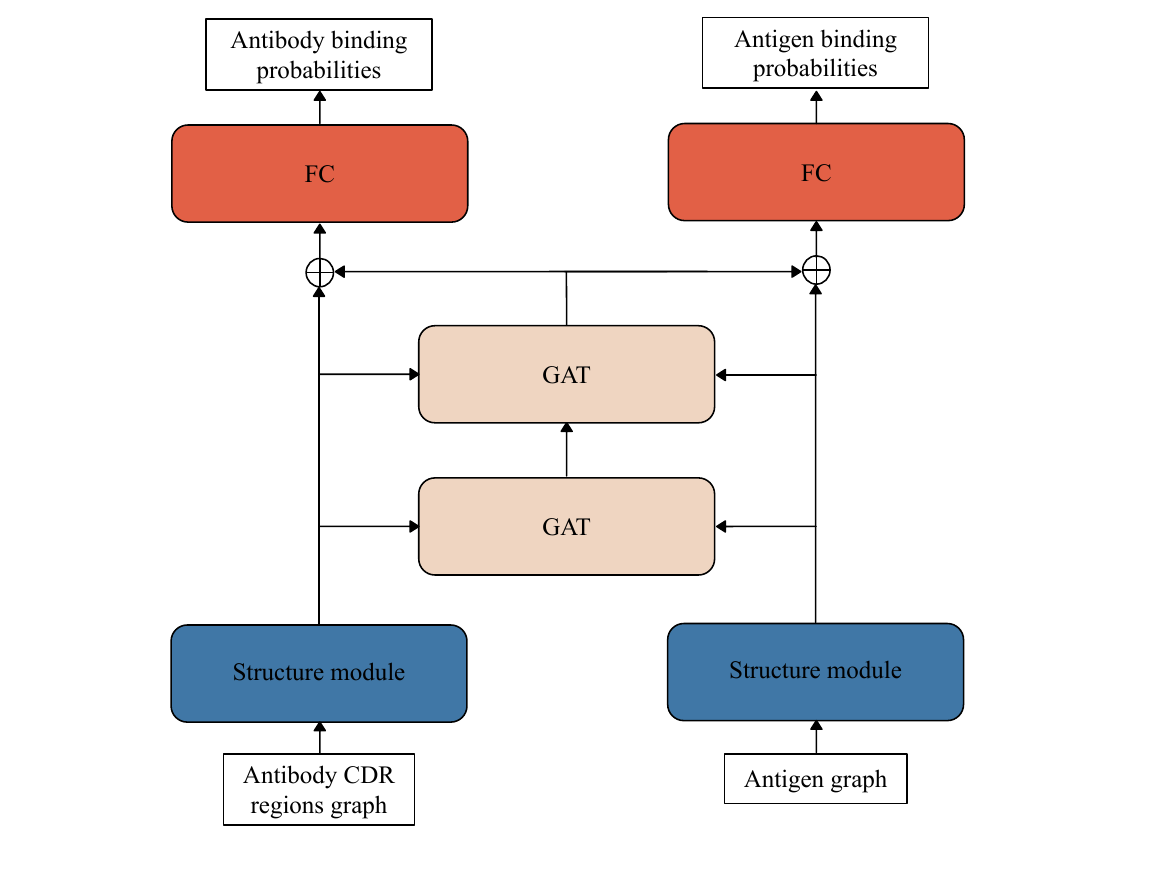}
\end{overpic}
\caption{\label{fig:IGEP} I-GEP models architecture.}
\end{figure}

\subsection{\label{sec:IGEP} I-GEP}
\begin{figure}[h]
\vspace{0cm}
\centering
\begin{overpic}
[trim=0cm 0cm 0cm 0cm,clip,width=0.99\linewidth, grid=false]{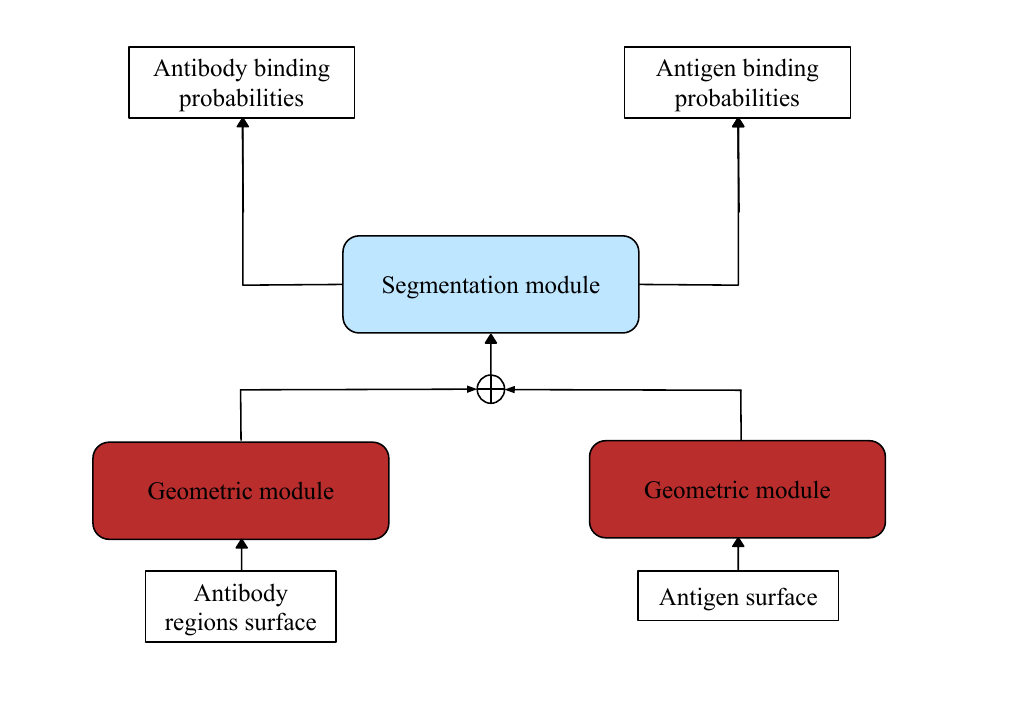}
\end{overpic}
\caption{\label{fig:OGEP} O-GEP models architecture.}
\end{figure}

Our I-GEP model is a method for predicting epitopes and paratopes using a graph-based approach that captures the inner structure of a protein. Each residue is represented as a node in a graph, and edges are created between the 15 closest neighbouring residues within 10 \AA. The I-GEP model has two main components: a structural module that computes an embedding for each residue using the graph structure and a graph attention network (GAT) that combines information from both the antigen and antibody residues. The network then predicts both epitope and paratope residues simultaneously using a fully connected layer, as shown in Fig. \ref{fig:OGEP}.

To improve the accuracy of our predictions, we integrate geometric information into the I-GEP model using two different approaches.
In the first approach, EPMP$_{xyz}$, we use graph convolutional network layers in the structural module as in EPMP \cite{del2021neural}, but we include the centred 3D coordinates of residues in the input features. 
The second approach, \textit{$E(n)$-EPMP}, uses the $E(n)$ invariant layer encoder from EGNN \cite{satorras2021n} instead of graph convolutional networks. This approach considers only the distances between residues, making it invariant to translations, rotations, and reflections on the residue positions in each molecule. 

\subsection{\label{sec:OGEP} O-GEP}
Our O-GEP model operates on the protein's surface and includes a geometric module that uses the surface's geometry to spread information across it. This process generates features that are then combined and shared between the antibody and antigen through fully connected layers (segmentation module), resulting in an interaction probability for each point on the surface, as shown in Fig. \ref{fig:OGEP}.

We explore two different models for the geometric module. As a baseline, we use PointNet \cite{qi2017pointnet} to recreate the architecture proposed in PiNet \cite{PiNet}. The second model employs diffusion layers from DiffNet \cite{sharp2022diffnet} to propagate features on the surface. This makes our model robust against surface perturbations and suitable for handling meshes and point clouds with fewer points.

We further examine the impact of using the Heat Kernel Signature (HKS) as an extra geometric descriptor input. The HKS \cite{sun2009hks} is a concise point-wise spectral signature which summarizes local and global information about the intrinsic geometry of a shape by capturing the properties of the heat diffusion process on the surface. 
One of the key benefits of using HKS is that it remains stable even under minor surface perturbations, thus enabling it to withstand even conformational rearrangements of the proteins. 
To utilize the HKS descriptor, we concatenate it with the input features at each point on the surface and then pass the concatenated data through the geometric module.

To transfer the binding probabilities from the protein's surface to the residues, we utilized the average of all the points on the surface that correspond to the same residues. This method ensures that the binding probabilities are accurately represented in the residue space, enabling us to make reliable predictions about epitope and paratope locations.

\subsection{Training}

To handle imbalanced binary classification tasks, the networks were trained using the class-weighted binary cross-entropy loss and the Adam SGD optimizer. 
For parameter tuning, we performed a hyperparameter search on the validation set. 
We train each model with five random seeds, and for each run, we keep the models' weights that performed the best on the validation set. 
During training, we also randomly rotate instances of the dataset to increase the robustness of the models.
See Appendix \ref{sec:hyperparams} for more details.

\subsection{Evaluation}

Given the significant disparity in class sizes, we utilize Matthew's correlation coefficient (MCC) between the residues' classification as our main benchmarking metric for model evaluation. 
We also report the area under the receiver operating characteristic curve (AUC ROC) and the area under the precision-recall curve (AUC PR) as used in \cite{PiNet,del2021neural}. 
All reported values are aggregated across five random seeds to ensure the robustness of our findings.

\section{Results}
In this section, we report the results of our experiments and demonstrate the contribution of geometric information on the task of epitope-paratope prediction.



\begin{table}[t]
\caption{\label{tab:graph_res} Results from I-GEP models}
\vskip 0.01in
\centering
\begin{sc}
\scriptsize
\setlength{\tabcolsep}{3pt}
\begin{tabular}{l c c c c}
 & MCC& AUC ROC& AUC PR\\ \cmidrule(lr){2-4}
 & \multicolumn{3}{c}{Antigen}\\ \toprule
 EPMP&$ 0.09 \pm 0.01$&$ 0.61 \pm 0.01$&$ 0.12 \pm 0.00$\\ \midrule[0.2pt]
EPMP$_{xyz}$&$ 0.10 \pm 0.01$&$ 0.63 \pm 0.01$&$ 0.15 \pm 0.01$ & \\ 
$E(n)$-EPMP&$ \mathbf{0.14 \pm 0.01}$&$ \mathbf{0.68 \pm 0.02}$&$ \mathbf{0.16 \pm 0.01}$\\ \midrule
 & \multicolumn{3}{c}{Antibody} \\  \toprule
 EPMP&$ 0.39 \pm 0.02$&$ 0.79 \pm 0.01$&$ 0.53 \pm 0.01$\\ \midrule[0.2pt]
EPMP$_{xyz}$&$ 0.38 \pm 0.02$&$ 0.79 \pm 0.01$&$ 0.53 \pm 0.01$\\ 
$E(n)$-EPMP&$ \mathbf{0.44 \pm 0.11}$&$ \mathbf{0.82 \pm 0.07}$&$ \mathbf{0.60 \pm 0.10}$\\\midrule
\end{tabular}
\end{sc}
\vskip -0.02in
\end{table}
\textbf{I-GEP results}
We conducted experiments to evaluate the effectiveness of incorporating geometric information by comparing our proposed models from Section \ref{sec:IGEP} with the EPMP model proposed in \cite{del2021neural}. Our results, presented in Table \ref{tab:graph_res}, clearly demonstrate that the inclusion of geometric information leads to a meaningful increase in performance. Specifically, the use of the $E(n)$ invariant layer ({\sc $E(n)$-EPMP}) resulted in an improvement in all metrics for both antibody and antigen.

\begin{table}[t]
\caption{\label{tab:surface_res}Results from O-GEP models. In addition to the physicochemical features, we test different combination of geometric information: 3d coordinates {\sc (xyz)} and Heat Kernel Signature {\sc (HKS)}. For the {\sc DiffNet} models, we consider both the point cloud ($_{pc}$) and the mesh ($_{m}$) of the surface.}
\vskip 0.00in
\centering
\begin{sc}
\scriptsize
\setlength{\tabcolsep}{3pt}
\begin{tabular}{l c c c}
 & MCC& AUC ROC& AUC PR\\ \cmidrule(lr){2-4}
 & \multicolumn{3}{c}{Antigen}\\ \toprule
 PiNet {\tiny(xyz)} &$ 0.39 \pm 0.05$&$ 0.89 \pm 0.01$&$ 0.44 \pm 0.02$\\
PiNet {\tiny(xyz+hks)}&$ 0.30 \pm 0.04$&$ 0.87 \pm 0.02$&$ 0.37 \pm 0.06$\\ \midrule
DiffNet$_{pc}$ {\tiny(xyz)}&$ 0.41 \pm 0.06$&$ \mathbf{0.90 \pm 0.01}$&$ 0.49 \pm 0.02$\\
DiffNet$_{pc}$ {\tiny(hks)}&$ 0.07 \pm 0.05$&$ 0.66 \pm 0.02$&$ 0.14 \pm 0.01$\\ 
DiffNet$_{pc}$ {\tiny(xyz+hks)} &$ \mathbf{0.44 \pm 0.03}$&$ \mathbf{0.90 \pm 0.01}$&$ \mathbf{0.50 \pm 0.02}$\\ \midrule
DiffNet$_{m}$ {\tiny(xyz)}&$ 0.42 \pm 0.03$&$ \mathbf{0.90 \pm 0.01}$&$ 0.48 \pm 0.05$\\
DiffNet$_{m}$ {\tiny(hks)}&$ 0.09 \pm 0.02$&$ 0.64 \pm 0.02$&$ 0.14 \pm 0.01$\\ 
DiffNet$_{m}$ {\tiny(xyz+hks)}&$ 0.42 \pm 0.06$&$ \mathbf{0.90 \pm 0.01}$&$ 0.46 \pm 0.07$\\\midrule
 & \multicolumn{3}{c}{Antibody} \\  \midrule
 PiNet {\tiny(xyz)}&$ 0.26 \pm 0.12$&$ 0.77 \pm 0.03$&$ 0.52 \pm 0.08$\\
PiNet {\tiny(xyz+hks)}&$ 0.22 \pm 0.05$&$ 0.74 \pm 0.00$&$ 0.47 \pm 0.02$\\ \midrule
DiffNet$_{pc}$ {\tiny(xyz)}&$ 0.30 \pm 0.06$&$ 0.79 \pm 0.01$&$ 0.56 \pm 0.03$\\ 
DiffNet$_{pc}$ {\tiny(hks)}&$ 0.44 \pm 0.03$&$ \mathbf{0.85 \pm 0.00}$&$ 0.68 \pm 0.01$\\
DiffNet$_{pc}$ {\tiny(xyz+hks)}&$ 0.23 \pm 0.06$&$ 0.77 \pm 0.04$&$ 0.51 \pm 0.05$\\\midrule
DiffNet$_{m}$ {\tiny(xyz)}&$ 0.24 \pm 0.08$&$ 0.78 \pm 0.02$&$ 0.52 \pm 0.03$\\
DiffNet$_{m}$ {\tiny(hks)}&$ \mathbf{0.49 \pm 0.01}$&$ \mathbf{0.85 \pm 0.00}$&$ \mathbf{0.69 \pm 0.01}$\\ 
DiffNet$_{m}$ {\tiny(xyz+hks)}&$ 0.28 \pm 0.06$&$ 0.77 \pm 0.02$&$ 0.52 \pm 0.04$\\ \midrule
\end{tabular}
\end{sc}
\end{table}
\textbf{O-GEP results}
To test the performance of O-GEP models, we consider the methods proposed in Section \ref{sec:OGEP} with different combinations of input features.
The results are summarized in Table \ref{tab:surface_res}. Incorporating diffusion layers ({\sc DiffNet}) along with 3D coordinates and Heat Kernel Signature as additional features consistently outperformed the baseline method {\sc PiNet}. 
The use of these techniques led to an MCC score twice as high as that obtained by the I-GEP models. However, unlike epitope prediction, the paratope prediction did not show the same level of improvement with O-GEP models. In this case, the best results were achieved by considering only the HKS features and diffusion layers.



\newcommand\wc{0.23\textwidth}
\begin{figure}[h]
\centering
\begin{subfigure}[b]{\wc}
\centering
\begin{overpic}
[trim=8cm 8cm 9.5cm 7cm,clip,width=0.65\linewidth, grid=false]{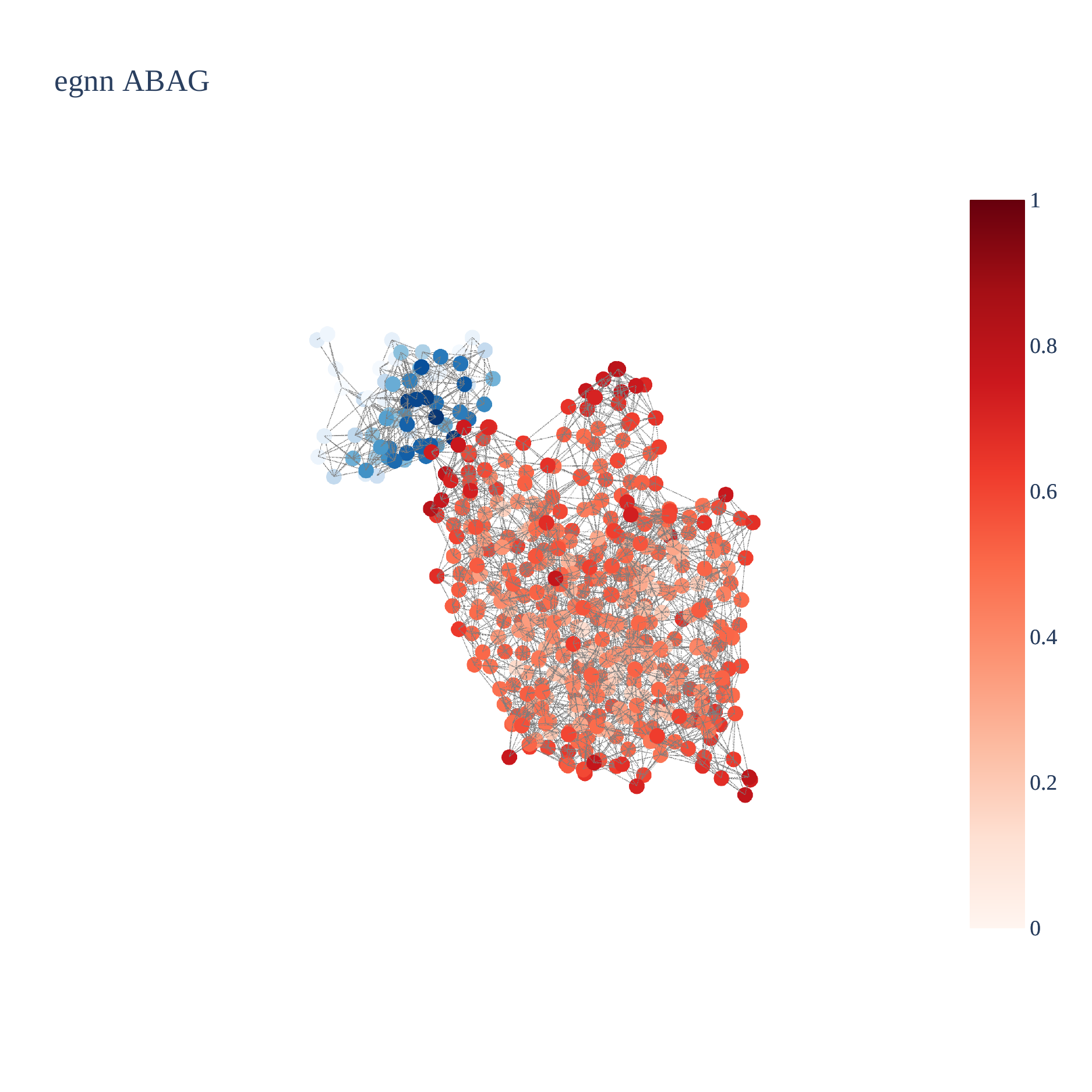}
\end{overpic}
\caption{\label{fig:graph_res} I-GEP}
\end{subfigure}
\hfill 
\begin{subfigure}[b]{\wc}
\begin{overpic}
[trim=12cm 9cm 11.5cm 10cm,clip,width=0.99\linewidth, grid=false]{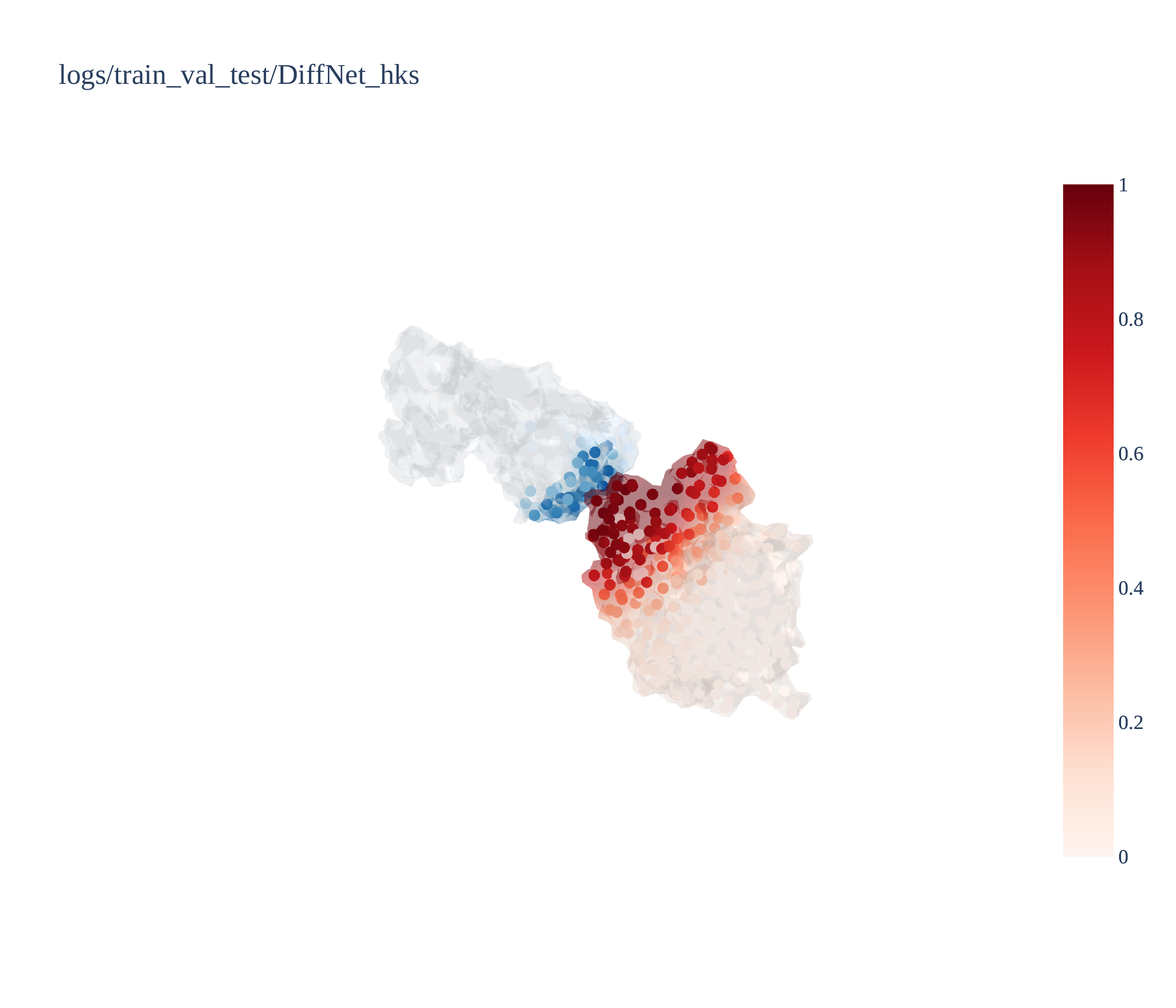}
\end{overpic}
\caption{\label{fig:surf_res} O-GEP}
\end{subfigure}
\caption{\label{fig:mol_res} Qualitative example on the 4jr9 pdb. We plot on the residuals the binding probability as increasing intensity colors: blue for the antibody and red for the antigen.}
\end{figure}

\textbf{Qualitative results} 
The qualitative examples shown in Figure \ref{fig:mol_res} clearly demonstrate the improved performance of O-GEP models over I-GEP.
Figure \ref{fig:graph_res} shows the results of the \textit{$E(n)$-EPMP} on the residual graph.  
The epitope prediction focuses on sparse regions of the antigene, such as the spiky edges. In contrast, paratope prediction concentrates on the residues closest to the antigen.
In Figure \ref{fig:surf_res}, the predictions of {\sc DiffNet$_{pc}$ {\tiny(xyz+hks)}} are shown on both the surface and residues of the molecules. The predictions are highly localized on the region nearest to the binding molecule.
It's worth noticing that the 3d coordinates given as input to the models are centred and randomly rotated, providing no prior knowledge of the binding region.

\section{Conclusions}


We investigated the effectiveness of geometric deep learning techniques in predicting antibody-antigen interactions. Our results indicate that incorporating geometric information is crucial for accurately predicting epitope and paratope regions. Specifically, the use of invariant representation in I-GEP models outperformed previous models, and O-GEP models with diffusion layers and additional geometric features achieved state-of-the-art performance. Our study highlights the potential of geometric deep learning in computational biology. Future research could explore using spectral shape analysis to address the more complex problem of conformational rearrangement in antigen-antibody binding \cite{stanfield1994major}.

\section*{Acknowledgements}
The authors would like to thank Joshua Pan and Karl Tuyls for their valuable feedback on the paper. C.D. was supported by The Elise Mobility Program funded from the European Union’s Horizon 2020 research and innovation programme under ELISE Grant Agreement No. 951847.
This work was supported by the ERC Grant no.802554 "SPECGEO" and PRIN 2020 project no.2020TA3K9N "LEGO.AI".

\medskip
\bibliographystyle{icml2023}
\bibliography{eg_bib}


\newpage
\appendix
\section{\label{sec:hyperparams} Hyper-parameters}

After the hyperparameter search, we found that the best learning rates were:
$10^{-3}$ for {\sc EPMP} and {\sc PiNet}, 
$10^{-2}$ for {\sc $E(n)$-EPMP}, 
$5*10^{-3}$ for {\sc DiffNet}.
We trained all the models for 200 epochs and kept the weights that performed the best on the validation metrics during training.

The surface generated by PyMOL are composed of around 14k points. To ease and fast the training procudere we subsampled the surface considering only 2k points. In the case of point clouds we usa a random subsampling during training, while for the mesh we used a simplification method base on quadric error metrics.

\subsection{Layer dimensions}
For the {\sc EPMP$_{xyz}$} model, we use a graph convolution layer with inner dimension 31 and two GAT layers with inner dimension 62. In contrast, for the {\sc $E(n)$-EPMP}, we use one $E(n)$-invariant layer with an inner dimension of 28 and two GAT layers with inner dimension 56.

For all the O-GEP models, the geometric module comprises two layers with dimensions 64 and 128, while the segmentation module is composed of two layers with dimensions 64 and 32.

\end{document}